\documentclass[twocolumn,aps,preprintnumbers,amsmath,amssymb,superscriptaddress]{revtex4}
\usepackage{subfigure,multirow}
\usepackage{longtable}

\usepackage{graphicx, color}
\usepackage{physics}
\usepackage{chngpage}

\def\bse{\begin{eqnarray*}}
\def\ese{\end{eqnarray*}}
\def\be{\begin{eqnarray}}
\def\ee{\end{eqnarray}}
\def\bq{\begin{equation}}
\def\eq{\end{equation}}

\begin{document}

\title{Dark matter signals from timing spectra at neutrino experiments}

\author{Bhaskar Dutta}
\email{dutta@physics.tamu.edu}
\affiliation{Mitchell Institute for Fundamental Physics and Astronomy, Department of Physics and Astronomy, Texas A\&M University, College Station, TX 77843, USA}
\author{Doojin Kim}
\email{doojin.kim@tamu.edu}
\affiliation{Mitchell Institute for Fundamental Physics and Astronomy, Department of Physics and Astronomy, Texas A\&M University, College Station, TX 77843, USA}
\affiliation{Department of Physics, University of Arizona, Tucson, AZ  85721  USA}
\author{Shu Liao}
\email{ikaros@physics.tamu.edu}
\affiliation{Mitchell Institute for Fundamental Physics and Astronomy, Department of Physics and Astronomy, Texas A\&M University, College Station, TX 77843, USA}
\author{Jong-Chul Park}
\email{jcpark@cnu.ac.kr}
\affiliation{Department of Physics, Chungnam National University, Daejeon 34134, Republic of Korea}
\author{Seodong Shin}
\email{sshin@jbnu.ac.kr}
\affiliation{Department of Physics \& IPAP, Yonsei University, Seoul 03722, Republic of Korea}
\affiliation{Center for Theoretical Physics of the Universe, Institute for Basic Science, Daejeon 34126, Repulic of Korea}
\affiliation{Department of Physics, Jeonbuk National University, Jeonju, Jeonbuk 54896, Republic of Korea}
\author{Louis E. Strigari}
\email{strigari@tamu.edu}
\affiliation{Mitchell Institute for Fundamental Physics and Astronomy, Department of Physics and Astronomy, Texas A\&M University, College Station, TX 77843, USA}

\preprint{
\begin{minipage}[b]{1\linewidth}
\begin{flushright}
MI-TH-1925
\end{flushright}
\end{minipage}
}

\begin{abstract}
We propose a {\it novel} strategy to search for new physics in timing spectra at low-energy neutrino experiments using a pulsed beam, envisioning the situation in which a new particle comes from the decay of its heavier partner with a finite particle width.
The timing distribution of events induced by the dark matter (DM) candidate particle scattering at the detector may populate in a relatively narrow range, forming a ``resonance-like'' shape. 
Due to this structural feature, the signal may be isolated from the backgrounds, in particular when the backgrounds are uniformly distributed in energy and time. For proof of the principle, we investigate the discovery potential for DM from the decay of a dark photon in the ongoing COHERENT experiment, and show the exciting prospects for exploring the associated parameter space with this experiment. We analyze the existing CsI detector data with a timing cut and an energy cut, and find, for the first time, an excess in the timing distribution which can be explained by such DM. We compare the sensitivity to the kinetic mixing parameter ($\epsilon)$ for current/future COHERENT experiments with the projected limits from LDMX and DUNE.
\end{abstract}
\maketitle

Numerous theoretical and experimental ideas have been put forth to identify the  mass and associated interactions of DM candidate particles.
Since traditional WIMP-based searches
have not yet detected DM~\cite{Aprile:2018dbl}, expanding the search of parameter space is well-justified~\cite{Battaglieri:2017aum}.
Many models of light DM ($\lesssim$ GeV) emerge from a hidden/visible sector where light mediators (e.g., a dark photon) interact with DM~\cite{Huh:2007zw, Pospelov:2007mp, Hooper:2008im, Cheung:2009qd, Essig:2010ye, Essig:2013lka, Dutta:2019fxn}.
Because the DM mass is light in these models, it is difficult to detect such DM in traditional WIMP-based direct detection experiments.

In this paper we develop a {\it novel} strategy to search for light DM which couples with light mediators and apply it to the data from the ongoing COHERENT experiment~\cite{Akimov:2018ghi}.
COHERENT makes use of a proton beam which impinges on a Hg target at the Spallation Neutron Source (SNS).
Among the produced pions, $\pi^+$ decays create prompt muon neutrinos and delayed anti-muon and electron neutrinos.
The measured energy spectra have been used to investigate new physics associated with neutrino non-standard interactions (NSI)~\cite{Ohlsson:2012kf,Miranda:2015dra} due to heavy or light mediators~\cite{Coloma:2017egw,Coloma:2017ncl,Liao:2017uzy,Dent:2017mpr,Billard:2018jnl, Lindner:2016wff, Farzan:2018gtr, Brdar:2018qqj}, generalized scalar and vector neutrino interactions~\cite{AristizabalSierra:2018eqm}, hidden sector models~\cite{Datta:2018xty}, and sterile neutrinos~\cite{Kosmas:2017zbh,Blanco:2019vyp}. It also sets independent constraints on the effective neutron size distribution of CsI~\cite{Ciuffoli:2018qem,AristizabalSierra:2019zmy,Papoulias:2019lfi}.
Since the proton beam is pulsed, the measured timing spectra may be used to distinguish between prompt and delayed events. The combined timing and energy spectra have been utilized to understand new physics models with neutrino flavor-dependent NSI~\cite{Dutta:2019eml}.

We show how both the timing and energy data from the COHERENT experiment can be used to search for light, $\lesssim 1$ GeV, DM.
The DM event under consideration is initiated by the production of a dark photon decaying into a pair of DM particles (e.g., Refs.~\cite{deNiverville:2015mwa,Ge:2017mcq}). 
A DM particle would then induce a nuclear recoil event at the detector.
The dark photon production can occur from both  $\pi^-$ and $\pi^0$.
Most of the $\pi^{-}$ are stopped inside the Hg target and can create a dark photon via the absorption process, $\pi^{-} + p \rightarrow n + A'$, followed by the decay of the dark photon $A'$ to a DM pair~\cite{deNiverville:2015mwa}.
The dark photon is emitted isotropically in this $\pi^{-}$ absorption process.
The $\pi^{0}$ may produce an ordinary photon and a $A^\prime$~\cite{Ge:2017mcq}. 
Since the $\pi^0$ move somewhat relativistically, the resulting DM lies relatively in the forward direction. 
Nevertheless, we find that the DM flux reaching to the COHERENT detectors, which are located $\sim 90^\circ$ from the beam direction~\cite{Akimov:2018ghi} (see Appendix), is comparable to that from the $\pi^-$ absorption.
Further, there are additional contributions from $\pi^{\pm} + p/n \rightarrow n/p + \pi^0$.

The method that we develop to search for DM utilizes both the energy and timing spectra of the DM-initiated nuclear recoil events.
We focus on the timing and energy spectra for the DM produced from the $\pi^{-}$ absorption and the $\pi^0$ decay. 
In the COHERENT experiment, the $\pi^{-}$ ($\pi^+$) and $\pi^0$ abundances per proton on target are 0.05 (0.11) and 0.1, respectively~\cite{deNiverville:2015mwa, akimov,Coherent}.
The produced $A^\prime$ is mostly relativistic unless its mass is $\sim$ 138 MeV.


We first derive the timing spectrum of DM-induced nuclear recoil events along with  their energy distribution, and then compare the DM case to that of Standard Model (SM) neutrinos.
The signal under consideration is initiated by production of a dark photon $A'$ from the decay of the $\pi^-$-$p$ mesic state and $\pi^0$ decay through kinetic mixing.
$A'$ production and its subsequent decay to DM $\chi$ are governed by the following interaction Lagrangian:
\begin{equation}
{\cal L}_{\rm int} \supset g_\chi
A'_\mu\bar{\chi}\gamma^\mu\chi+e_q \epsilon_1^q A'_\mu{\bar q}\gamma^\mu q \,, \label{eq:lag}
\end{equation}
where $e_q=eQ_q$, $g_\chi$ and $\epsilon_1^q$ are dark-sector gauge coupling and kinetic mixing parameter (associated with the mixing between the $\gamma$ and new gauge Boson ${\epsilon\over{2}} F^{\mu\nu\prime} F_{\mu\nu}$~\cite{Holdom:1985ag,delAguila:1988jz,Babu:1997st}), respectively.
This generic-looking Lagrangian can be accommodated in the context of a model, e.g., ~\cite{deNiverville:2015mwa,Dutta:2019fxn}.

Let us suppose that $A'$ is produced at $t_F$ where $t_F$ is the timing of $\pi^{-,0}$ production induced by the 1 GeV SNS beam which is $0.6~\mu$s wide and pulsed at 60 Hz.
We then assume that $A'$ flies for $v_{A'}(t-t_F)$ along the $\theta$ direction with respect to the line joining the Hg target and the detector (see Appendix), and decays to a $\chi$ pair.
One of the $\chi$'s then may travel towards the detector for $v_\chi t'$.
Denoting the timing measured at the detector by $T$, we see that $T$ is the sum of $t$ and $t'$, i.e., $T= t+t'(v_{A'}(t-t_F), t-t_F, \cos\theta)$, where we explicitly express $t'$ as a function of $t-t_F$ and $\cos\theta$.
We are interested in the differential number of events in $T$, or equivalently the DM flux at the detector of interest, $f(T)=dN_\chi/dT$. 
Parameterizing the angular distribution of dark photons by $g(\cos\theta)$, we find
\begin{equation}
\frac{d^2N_{A'}}{dt d\cos\theta}= g(\cos\theta) \cdot \frac{1}{\tau_{A'}}e^{-\frac{t-t_F}{\tau_{A'}}}\Theta(t-t_F)\,,
\end{equation}
where $\Theta(x)$ is the step function. 
$g(\cos\theta)$ is 1/2 for dark photons from the $\pi^-$ absorption.
We then obtain
\begin{equation}
f(T) \propto \int d\cos\theta \left| \frac{dT}{dt} \right|^{-1}\frac{d^2N_{A'}}{dt d\cos\theta}\,.
\end{equation}
A simple geometry consideration gives 
\begin{equation}
T = t+\frac{1}{v_\chi}\sqrt{x_0^2+v_{A'}^2(t-t_F)^2-2x_0 v_{A'}(t-t_F)\cos\theta}
\end{equation}
with $x_0$ being the distance between the Hg target and the detector. 
We consider both $\pi^-$ and $\pi^0$ contributions. 
In our calculation, we use the GEANT4~\cite{Agostinelli:2002hh} simulations for the COHERENT geometry to determine the angular and energy spectra of photons from pion absorption and decays~\cite{Coherent}.
We find that the $\pi^0$ contribution is bigger than the $\pi^-$ absorption.

\begin{figure}
\centering
    \includegraphics[width=8.4cm]{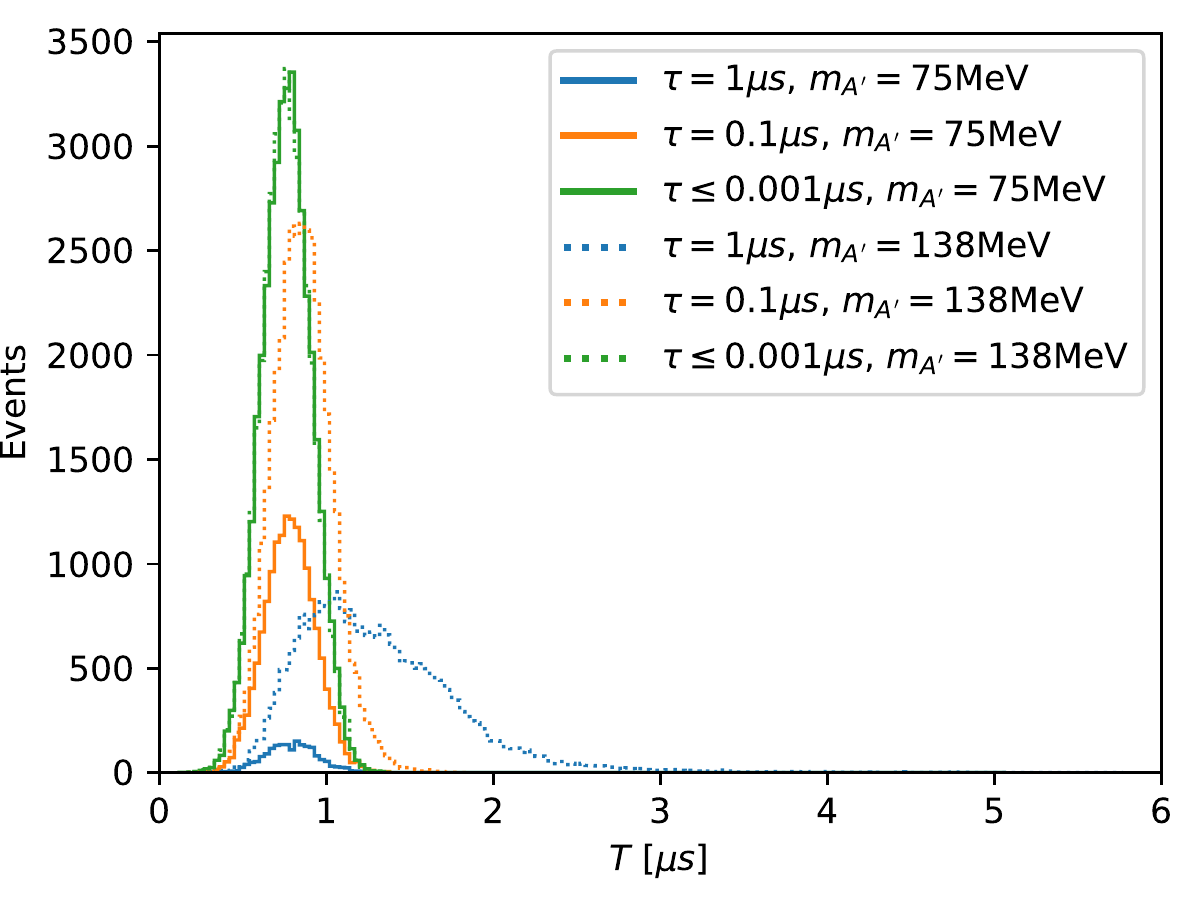}
    \includegraphics[width=8.4cm]{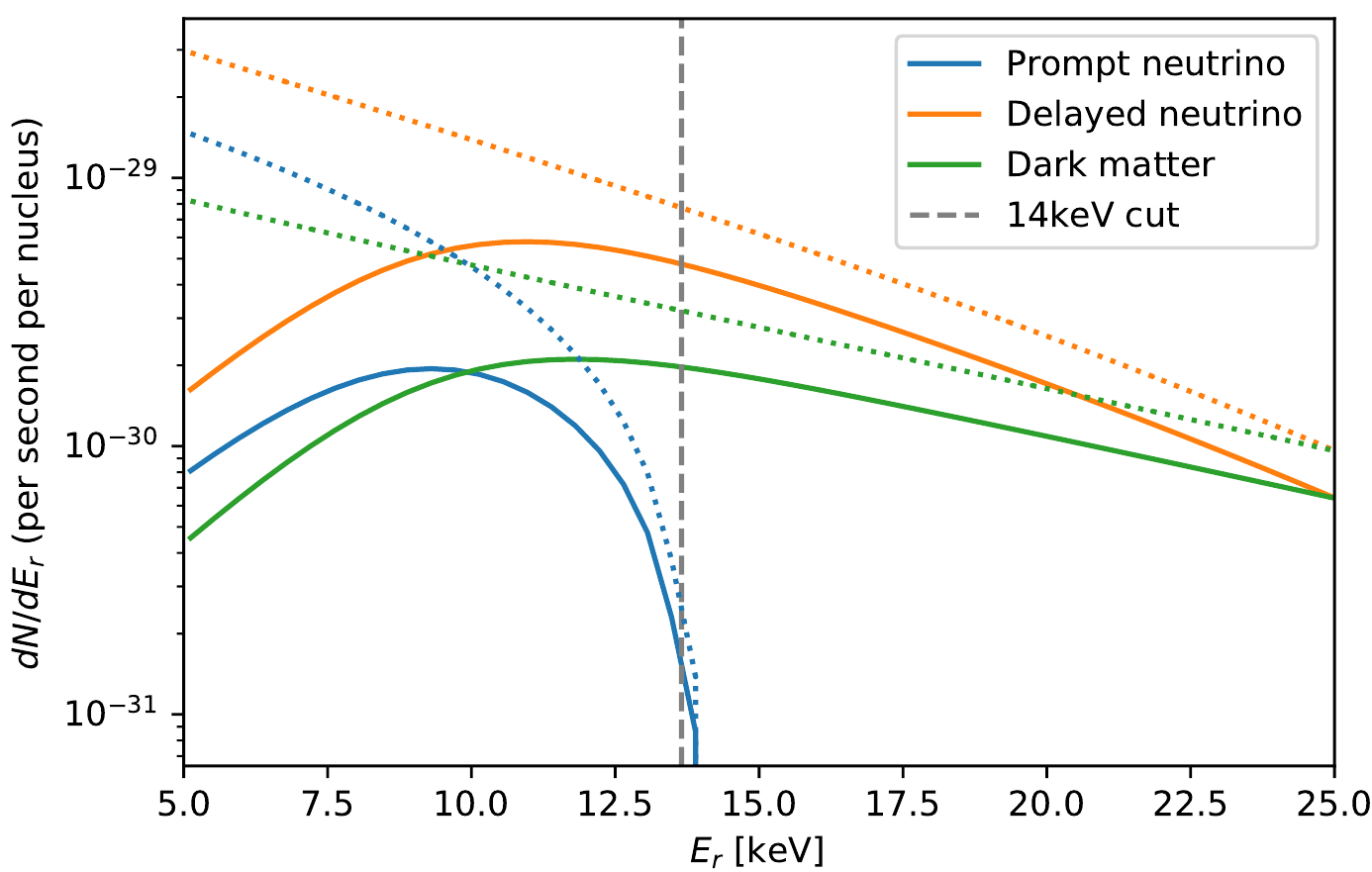}
    \caption{Top: Timing spectra of DM signal with three different values for $\tau_{A'}$, in a relativistic $A'$ scenario (solid) and a non-relativistic $A'$ scenario (dashed).
    Bottom: Nuclear recoil spectrum produced from neutrino and DM interactions with (solid) and without (dashed) experimental efficiencies.
    The vertical dashed line indicates the energy cut that is used to eliminate prompt $\nu$-induced events.
    }\label{fig:timing}
\end{figure}

The top panel of figure~\ref{fig:timing} demonstrates example timing spectra for a CsI detector, with three different choices for the rest-frame mean lifetime of $A'$.
The solid and dashed histograms are for a relativistic dark photon ($m_{A'}=75$ MeV) and a non-relativistic dark photon ($m_{A'}=138$ MeV), respectively, with $m_\chi$ fixed to 5 MeV.
Here the $\pi^-$ flux -- which is approximated by a Gaussian distribution with a mean value of 0.7 $\mu$s and a width of 0.15 $\mu$s to model the arriving time of the proton on target which  reproduces the timing spectrum in~\cite{akimov} -- is convoluted.
For the non-relativistic case, most of the $\chi$'s can reach the detector (modulo a factor of $(4\pi x_0^2)^{-1}$).
Not surprisingly, as $A'$ is shorter-lived, the spectrum width gets narrower, manifesting in a resonance-like bump feature more visibly.
By contrast, for the relativistic case, if $A'$ is long-lived, it decays far away from the detector so that only a small fraction of the $\chi$'s can reach the detector, contributing to the upper tail of the spectrum.
Therefore, relatively short-lived $A'$ would give more statistics.
Indeed, we see that most of DM events populate within $\sim 1.5~\mu$s which roughly corresponds to the mean value plus the width of the beam pulse.
Note that prompt neutrinos leave events within $\sim 1.5~\mu$s whereas delayed neutrinos spread out over a broad range~\cite{Akimov:2018ghi, akimov}.
So, requiring $T\lesssim 1.5~\mu$s essentially rejects most of delayed neutrino events while a large portion of prompt neutrino events and relativistic (non-relativistic) $A'$-induced DM events irrespective of $\tau_{A'}$ (with $\tau_{A'}\lesssim 0.1~\mu$s) are kept.

Regarding DM-nucleus scattering, we remark that in principle DM scattering can be governed by physics different from that for dark photon production encoded in Eq.~\eqref{eq:lag}.
Introducing a generic mediator of mass $M'$, DM-mediator coupling $g_D$, and quark-mediator coupling $e_q \epsilon_2^q$, we find that the differential spectrum in recoil energy $E_r$ of the target nucleus can be expressed as
\begin{eqnarray}
\frac{d\sigma}{dE_r} &=& \frac{e^2 (\epsilon_2^q)^2 g_D^2 Z^2\cdot |F(2m_NE_r)|^2}{4\pi p_\chi^2 (2m_NE_r+M^{'2})^2} \\
& &\times\left\{2E_\chi^2m_N\left(1-\frac{E_r}{E_\chi}-\frac{m_N E_r}{2E_\chi^2}\right)+E_r^2m_N\right\},\nonumber
\end{eqnarray}
where $F$ denotes the form factor and where $Z$ and $m_N$ are the atomic number and the mass of the target nucleus.
The underlying interaction is of dark-photon type for illustration.
We neglected $m_\chi$ in the curly brackets as $m_N \gg m_\chi$.
Clearly, the spectral behavior is (nearly) independent of $m_\chi$.
The bottom panel of figure~\ref{fig:timing} displays the expected nuclear recoil spectrum for $(M',m_\chi)=(75,5)$ MeV (green).
For comparison we show the $E_r$ distributions of the prompt neutrinos (blue) and the delayed neutrinos (orange) with (solid) and without (dashed) experimental efficiencies.
We see that prompt neutrino events occur almost entirely in the region $E_r \lesssim 14$ keV, so employing a lower cut at this energy removes the remaining prompt neutrinos, while retaining a large portion of the DM candidate events.


In order to analyze the COHERENT data using both the energy and timing spectra~\cite{akimov} from neutrinos and DM, we adopt the statistical method described in Ref.~\cite{Dutta:2019eml}. We allow for Poisson fluctuations of the background in each energy and time bin (model (c) of Ref.~\cite{Dutta:2019eml}), and fix the size of the neutron distribution to $R_n = 4.7$ fm. We also quote our results for $R_n=5.5$ fm which is the model independent central value obtained from the fit to the COHERENT data~\cite{Cadeddu:2017etk}. We examine two limiting cases:
i) the specific part of the energy and timing data in which the DM signal is predicted to appear, after removing as many neutrino-induced events as possible, and ii) the full energy and timing data.

As discussed in the previous section, we apply cuts $E_r>14$~keV (16 photoelectrons) and $T<1.5~\mu$s to substantially suppress both prompt and delayed neutrino events, but keep the DM candidate events~\footnote{Unless $m_{A'} \approx 138$~MeV and $\tau_{A'} \gtrsim 0.03~\mu$s.} for the published COHERENT data~\cite{akimov}. 
We also apply an upper-cut $E_r < 26$~keV  since the background is well understood for COHERENT in the range  5 to 26 keV~\cite{akimov}. 
The experimental efficiency is also given in~\cite{akimov}. 
After these cuts, we find 97 total events.
Out of them 49 events have been classified as the steady-state (SS) background, while 19 may be identified as delayed neutrino events forming the SM (i.e., neutrino) background.
There are also 3 events in the cut window arising from beam related neutron (BRN) backgrounds.
There is then an ``excess'' of 26 events which corresponds to a 2.4$\sigma$ statistical uncertainty. For $R_n=5.5$ fm, the significance becomes $\sim$3$\sigma$.
For calculating the significance we apply~\cite{Scholberg:2018vwg} 
\begin{equation}
\hbox{Excess}=\frac{{\rm signal}-{\rm SS}-{\rm BRN}-{\rm SM}}{\sqrt{2{\rm SS}+{\rm BRN}+{\rm SM}}}.
\end{equation}
We also calculate the significance from the likelihood ratio test for the DM fit to the excess and find the significance to be 1.98. 
The SS and BRN backgrounds emerge from measuring beam-on anti-coincident events and the simulation from GEANT4, respectively. 
We use the same systematic uncertainty $\sim 28\%$~\cite{Akimov:2017ade} which  incorporates flux, form factor, quenching factor and the signal acceptance uncertainties.

\begin{figure}
\centering
 \includegraphics[width=8.4cm]{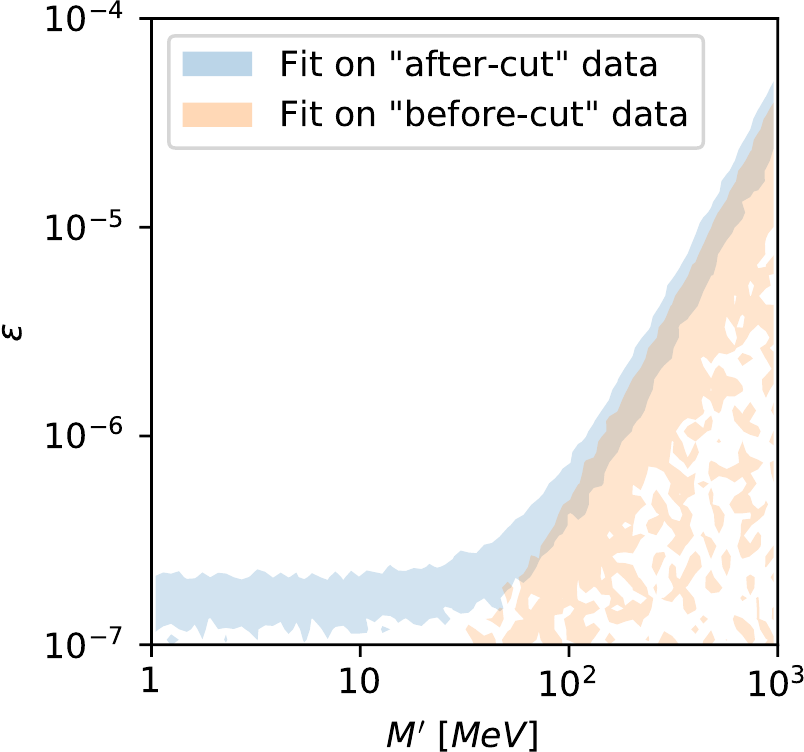}
 \includegraphics[width=8.4cm]{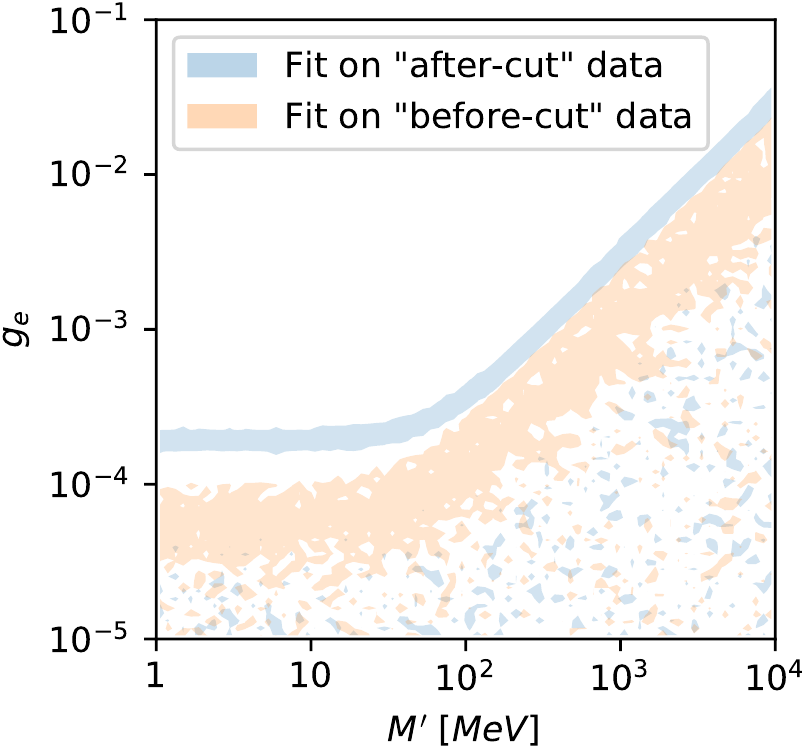}
    \caption{$1\sigma$ best fits to the ``before-cut'' data (orange) and the ``after-cut'' data (blue) for a DM interpretation (left panel) and a neutrino NSI interpretation (right panel). }
    \label{fig:combine}
\end{figure}

We first attempt to explain the excess with a DM hypothesis, again assuming that the DM scattering is governed by a different mediator.
We fit the selected events, varying the associated (effective) coupling constant $\epsilon$ and mediator mass $M'$ which is responsible for the interaction between the DM and the nucleus.
The left panel of figure~\ref{fig:combine} shows $1\sigma$-best fits to the data set with the cuts implemented (blue band).  The scattered ``islands" come from our MCMC sampling methods, with limited live points which cause some ``islands" in the result.
For comparison, the orange band shows the parameter space when performing a fit to the full energy and timing data at $1\sigma$.
We see that there exists an overlapping region, and further find that both ``before-cut'' and ``after-cut'' data sets are well accommodated by the parameter points with $M' \gtrsim 100$~MeV.
For comparison to the DM case, we determine whether a NSI neutrino hypothesis is able to fit both the before-cut and after-cut data.
For the neutrino case we consider a non-zero coupling $g_e$, the NSI in the $\nu_e$ neutral-current interaction.
As shown in the right panel of figure~\ref{fig:combine}, it is not possible to simultaneously fit both the before-cut and after-cut data sets with this neutrino hypothesis.
In fact, this NSI model does not show a good fit for the excess in the prompt timing bin (i.e., $T< 1.5~\mu$s). The fuzzy region at low $g_e$ shows at there is some statistical consistency with the SM in this region, in particular for the before-cut data. The situation becomes even worse with $g_\mu\neq0$, since it affects not only the delayed but also the prompt spectrum.

In the DM case, the parameter $\epsilon$  is defined as $\epsilon=\epsilon^q_1 \epsilon^q_2 \epsilon_D \sqrt{{\rm BR}_{A'\rightarrow \chi\chi}}$, where $\epsilon^q_1$ is the $q$-$A'$ kinetic mixing which describes the dark photon production from the $\pi^{-}$ absorption, $\epsilon^q_2$ is the quark-mediator kinetic mixing for the DM-nucleus scattering cross-section, and $g_D=e\epsilon_D$ is the DM-mediator coupling.
This is the most general description, since in a realistic model there can be more than one mediator, e.g., scalar and gauge boson mediators commonly occur in models with spontaneous symmetry breaking.
Of course, the best-fit contour can also be interpreted in the case where there exists only a single mediator, i.e., $M'=m_{A'}$.

The parameter choices that we use to obtain the best-fit points are $\tau_{A'}=1$~ns, $m_{A'}=75$~MeV and $m_\chi=5$~MeV.
However, we find that the best-fit points do not change in the $\epsilon-M'$ plane for the following variations:
i) for $\tau_{A'}\lesssim4$~ns, since the DM flux maximizes for $\tau_{A'}\lesssim 4$~ns with $m_{A'}<138$~MeV,
ii) for the non-relativistic case, i.e., $m_{A'}=138$~MeV, with $\tau_{A'}\lesssim 30$~ns and
iii) for any $m_\chi$ smaller than $m_{A'}/2$.
For relativistic scenarios with large $\tau_{A'}~(\geq 4~{\rm ns})$, the best-fit regions get scaled by the appropriate associated DM flux (see the top panel of figure~\ref{fig:timing}).
For non-relativistic scenarios with large $\tau_{A'}~(\geq 30~{\rm ns})$, it is not possible to fit before-cut and after-cut data sets simultaneously because DM will contribute to both before $1\mu$s and after $1\mu$s events. 
Figure~\ref{fig:combine} is shown for $R_n=4.7$ fm. However, the best fit contours do not change for $R_n=5.5$ fm.

Based on the above discussions, we describe the best-fit parameters for the following two scenarios.
\begin{itemize}
\item {\bf Single-mediator scenario}: In this case, $\epsilon^q \equiv \epsilon_1^q = \epsilon_2^q$ and the dark photon $A'$ should decay fast. Otherwise, $\epsilon$ is small, meaning that the DM-nucleus scattering is so small that a very small number of events would occur.
Here $\epsilon=\epsilon^q_1 \epsilon^q_2  \epsilon_D \sqrt{{\rm BR}_{A'\to \chi\chi}}\to({\epsilon^q})^{2}\epsilon_D \sqrt{{\rm BR}_{A'\to \chi\chi}}$.
We can choose $\epsilon_D=1/e$ to make $g_D=1$ which makes $\tau_{A'}$ small, and we can still make use of the left panel of figure~\ref{fig:combine} (where $\tau_{A'}$ is set to be $\leq 1$~ns).
Table~\ref{scenario1} shows the best $\epsilon^q$ for a few $M'(=m_{A'})$ values for which the resulting branching ratios for $\pi^0\to \gamma A'$ and $\pi^\pm\to e^\pm \nu A'$ agree with current precision data of $\pi^0$ and $\pi^\pm$~\cite{Tanabashi:2018oca}.
We do not report any numbers below $M'=50$~MeV as we find that the best-fit region with the before-cut data does not overlap with that with the after-cut data.
\begin{table}[h]
\centering
\begin{tabular}{ c|c|c|c|c}
\hline
\hline
~$M'$ &50& 75 & 100&1000\\\hline
$\epsilon^q$&~$3.5\times 10^{-4}$~~&~$4.4\times 10^{-4}$~~&~$5.5\times 10^{-4}$~~&~$4.6\times 10^{-3}$~~\\\hline\hline
\end{tabular}
\caption{Best-fit $\epsilon^q$ for a few $M'$ values (in MeV) for the single-mediator scenario.
}
\label{scenario1}\end{table}

\item {\bf Multi-mediator scenario}: Unlike the previous scenario, $\tau_{A'}$ is not necessarily small, since $\chi$ scatters off the target nucleus via a new mediator with large coupling  while the dark photon can decay to a pair of DM particles with a longer lifetime. 
    Table~\ref{scenario1} for a  single mediator scenario still holds with $\epsilon^q$ identified as $\sqrt{\epsilon^q_1\epsilon^q_2 \epsilon_D e}$. 
\end{itemize}

\begin{figure}
\centering
 \includegraphics[width=8.4cm]{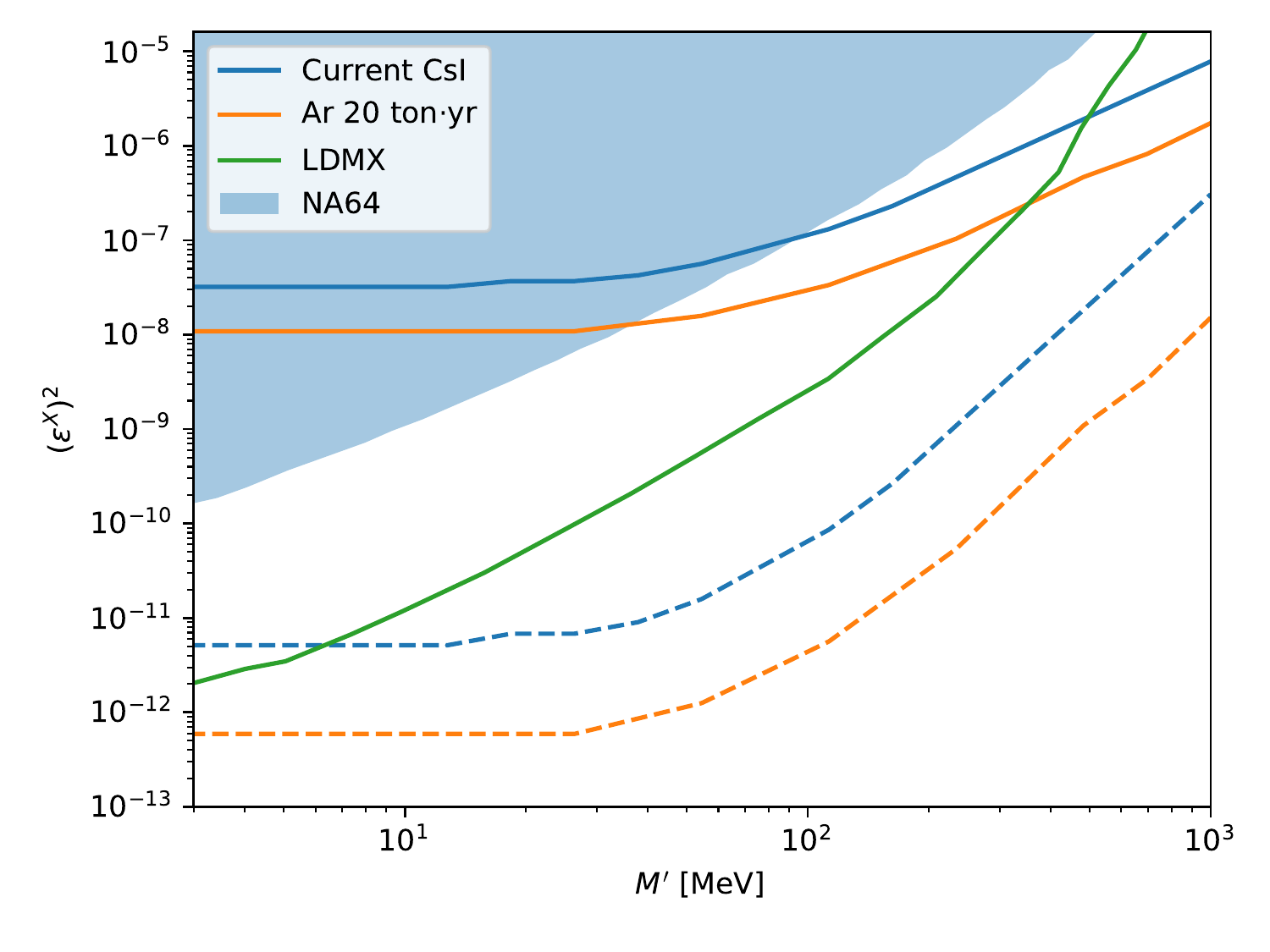}
    \caption{The coupling $(\epsilon^X)^2$ for mediator-nucleus coupling is shown as a function of $M^\prime$. 
    The solid (dashed) lines assume $\epsilon^X=\epsilon_1^q=\epsilon_2^q$ ($\epsilon^X=\epsilon_{1,2}^q$ with $\epsilon_{2,1}^q=10^{-2}$). $\epsilon^X$ for LDMX can be understood as $\epsilon$ in Ref.~\cite{Akesson:2018vlm}.}
    \label{fig:LDMX}
\end{figure}

The values of $\epsilon^q$ shown in Table~\ref{scenario1} are obtained assuming that the dark photon couplings to up and down quarks are proportional to their charges.
If, however, we want to use the universal charge (e.g., 1), then we need to scale the $\epsilon^q$ by $\sqrt{2Z/(9 A)}$ where $Z=54$ and $A=130$ for CsI.
The best-fit values of $\epsilon^q$ are below any existing bounds~\cite{Essig:2013vha,Dror:2017nsg} arising from meson decays, e.g., $K\rightarrow\pi+$invisibles~\cite{Artamonov:2009sz}. 
The model details become important for this constraint, i.e., whether it contains fully conserved current, additional Higgs sector~\cite{Davoudiasl:2014kua} and the value of $g_D$, etc. The excess can be  explained  in the allowed regions of parameter space of the realistic models, e.g., $U(1)_{T3R}$, $U(1)_{B-L}$,  etc.~\cite{deNiverville:2015mwa,Dutta:2019fxn}.
For example, for 50 MeV dark photon, the coupling ($e \epsilon$)$\sim 10^{-4}$, needed to explain the excess,  is well allowed by all the existing data in a generic vector-portal DM model~\cite{deNiverville:2015mwa}.
The COHERENT limit for NSI of neutrinos is better than any existing limit from various experiments using the timing plus energy data where the SM backgrounds cannot be sufficiently suppressed~\cite{Dutta:2019eml}.
However, for the DM analysis, since we have vetoed the SM neutrino backgrounds using the energy and timing cuts, we can obtain an even better reach in terms of new physics coupling.
Now assuming no excess above the backgrounds (which could be caused by an improper estimation of the time of creation of the pions at the target~\footnote{The COHERENT collaboration is investigating this possibility [Kate Scholberg, private communication].}),
the values of $\epsilon$  become smaller by a factor of 1.5 compared to those in Table~\ref{scenario1}, as shown in figure~\ref{fig:LDMX}. 

The future LDMX experiment~\cite{Akesson:2018vlm} will investigate the sub-GeV DM parameter space which arises from a dark photon decaying to DM, using an electron beam dump. 
We note that this parameter space is already being probed via nuclear recoils at COHERENT, therefore representing a complementary approach. 
In figure~\ref{fig:LDMX}, we compare the reach of $(\epsilon^X)^2$ as a function of mediator mass for the current COHERENT data  and for a future argon detector with the LDMX reach assuming that $\epsilon^q=\epsilon^e$. We also show the  existing limits from NA64~\cite{NA64:2019imj} relevant to both single- and multi-mediator scenarios in the plot assuming $\epsilon^q=\epsilon^e$.

Our current and projected limits are derived using the formalism of Ref.~\cite{Datta:2018xty}, and they are essentially governed by the $\pi^0$ contribution.
We show two scenarios: (i) the dark photon coupling ($\epsilon^q_1$) is the same as the mediator-nucleus coupling ($\epsilon^q_2$) and (ii) $\epsilon^q_1$ is fixed at $10^{-2}$ (current experimental constraint~\cite{Artamonov:2009sz}) with $\alpha_D\equiv g_D^2/(4 \pi)=0.5$.
We use a dark photon mass $m_{A^\prime}=75~\mathrm{MeV}$ and a DM mass $m_\chi=5~\mathrm{MeV}$. 
The figure, however, is unchanged for $m_{A'}\leq 138$ MeV, $m_\chi\le m_{A'}/2$ and $\tau_{A'}\leq 4$~ns. 
We also note that the reach of the current COHERENT data in probing $(\epsilon^X)^2$ in figure~\ref{fig:LDMX} is competitive with  DUNE experiment reach~\cite{DeRomeri:2019kic}.

In conclusion, we have argued that the timing information available in neutrino experiments with pulsed beam such as the COHERENT data is a powerful probe of new physics.
We have shown that the combination of energy and timing cuts can eliminate SM neutrino events very efficiently, thereby allowing the possibility of isolating DM-induced events.
As applied to the published COHERENT data, we find a considerable number of excess events over the expected backgrounds.
This excess of events may be explained by a dark matter hypothesis, and is unlikely to be explained by SM neutrino interactions.
We note that this conclusion is distinct from the results presented in Ref.~\cite{Dutta:2019eml}, who showed that using the full energy and timing data without using the cuts, a neutrino model is able to explain the data.
Even though we have presented a DM interpretation of the COHERENT data, it remains possible that the events may be explained by an unidentified background, by a systematic uncertainty on the observed steady-state background or by exotic beyond the SM scenarios. 
Distinguishing a background hypothesis from a DM hypothesis may be possible with timing and energy information on individual nuclear recoil events.
Our analysis strategy can be used to understand dark photon decaying to DM in similar COHERENT type set-ups with timing measurements~\footnote{As far as the beam pulse duration is less than $\sim 1~\mu$s to populate DM signals in the prompt timing bins and the pulse interval is greater than $\sim 10~\mu$s to sufficiently attenuate the flux of delayed neutrinos.}, e.g.,  JSNS$^2$~\cite{Ajimura:2017fld}, CCM~\cite{CCM}.

\section*{Acknowledgments}
We thank Phil Barbeau, Pilar Coloma, Yuri Efremenko, Pedro Machado, Grayson Rich, and Kate Scholberg for useful discussions.
BD, DK and LES acknowledge support from DOE Grant de-sc0010813.
The work of DK was supported in part by the Department of Energy under Grant No. DE-FG02-13ER41976/DE-SC0009913.
SL acknowledges support from COS-STRP (TAMU).
The work of JCP is supported by the National Research Foundation of Korea (NRF-2019R1C1C1005073 and NRF-2018R1A4A1025334).
The work of SS is supported by the National Research Foundation of Korea
(NRF-2017R1D1A1B03032076 and in partial by NRF2018R1A4A1025334).
This work was supported by IBS under the project code, IBS-R018-D1 (SS).

\section*{Appendix}

\begin{figure}[t]
    \centering
    \includegraphics[width=6.4cm]{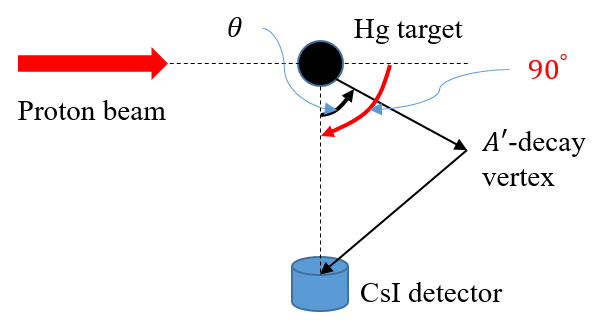}
    \caption{The experimental configuration for the CsI detector. Dark photon $A'$ exiting the Hg target flies in the $\theta$ direction with respect to the axis defined by the target and the detector. }
    \label{fig:csi}
\end{figure}

\noindent {\bf Brief description of COHERENT experiment}: The SNS in Oak Ridge National Laboratory produces neutrons using a 1.4\,MW, 1 GeV proton beam pulsed at 60 Hz on a mercury (Hg) target. 
A typical proton beam trace has a full-width at half-maximum (full-width) of 360 ns ($\sim 600$ ns). 
Various detectors are placed in the neutrino alley (see Fig. 10 of Ref.~\cite{Akimov:2018ghi} for their exact locations). 
For example, in Fig.~\ref{fig:csi}, we show the experimental configuration with the CsI detector at $90^\circ$ with respect to the beam direction. The dark photon flies at direction $\theta$ and decays to dark matter candidates which are then detected at the detector.

\bigskip

\begin{figure}[t]
    \includegraphics[width=8.4cm]{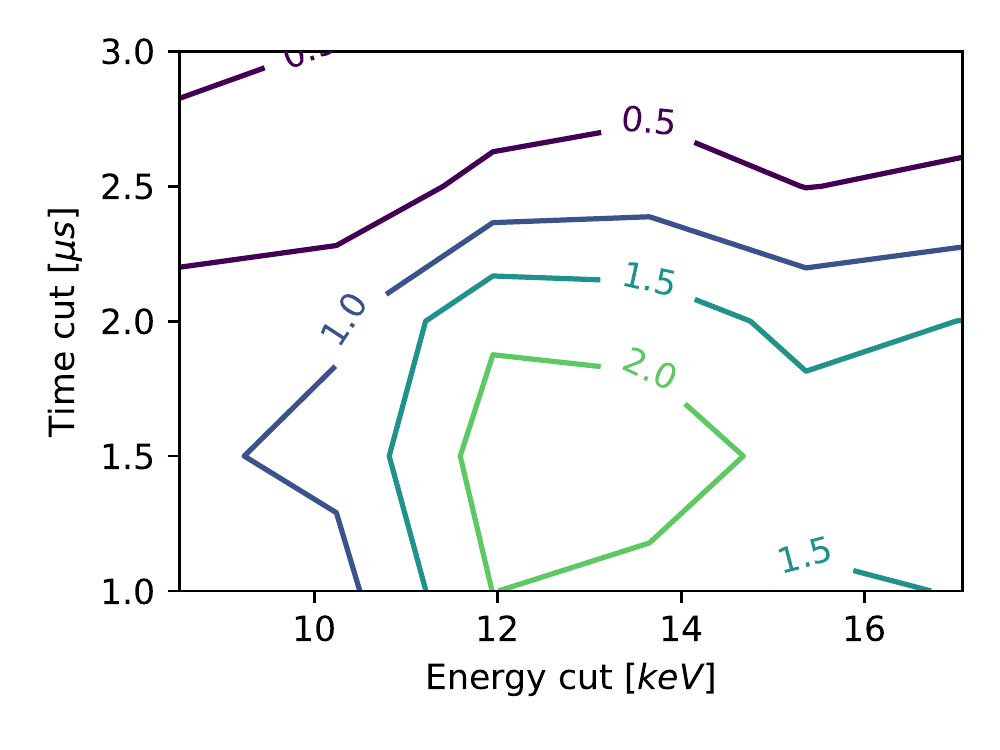}
    \caption{Statistical significance of  excess in the COHERENT data over the SM and other background (beam related neutron and steady state) is shown for various energy and timing cuts.
    }
    \label{fig:significance}
\end{figure}

\noindent {\bf Data analysis}: In Fig.~\ref{fig:significance}, we show the significance of  excess in the COHERENT data~\cite{akimov} over the SM and other background (beam related neutron and steady state) varying energy and timing cuts. 
We find the significance of the excess is maximized for $T<1.5\,\mu$s and $E_r>12$-$14$ keV. We use this set of cuts for our analysis.

Now, the region of $T<1.5\,\mu$s is preferred by dark matter events as we see from Fig. 1 (top panel) of the main text  and the neutrinos emerging from the pion decay ($\pi^+\rightarrow\mu^++\nu_\mu$) in the SM (but not the delayed neutrinos emerging from the $\mu$ decay with a lifetime being 2.2 $\mu$s), while our selected  region with $E_r>14$ keV and $T< 1.5 \,\mu$s is only preferred by the DM candidate particles since the DM candidates have more energy compared to the $\nu_\mu$ emerging from the pion decay ($<14$ keV).

\begin{figure}[t]
    \includegraphics[width=8.4cm]{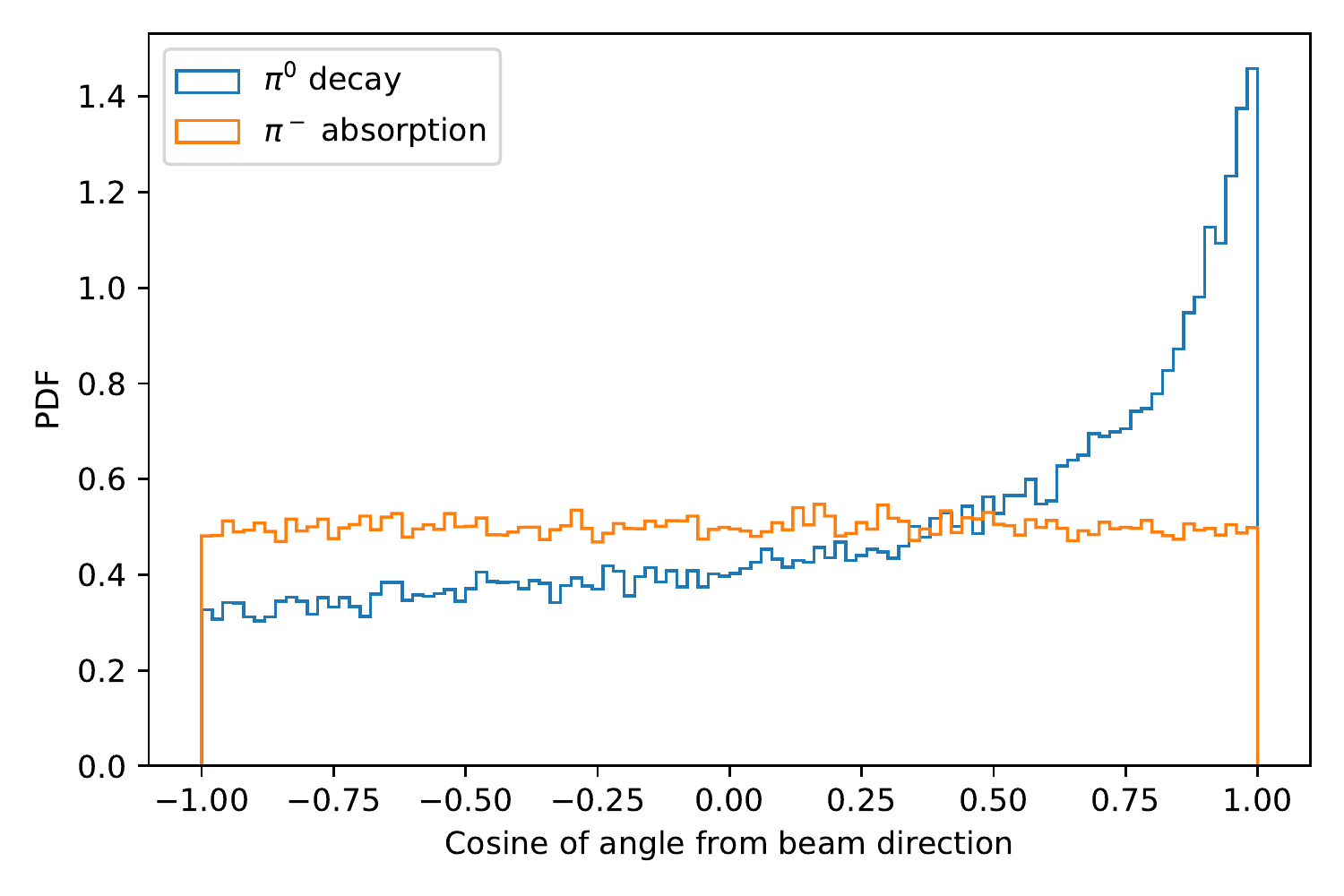}
    \caption{Unit-normalized angular distributions of dark photons emerging from $\pi^-$ absorption and $\pi^0$ decays.
    }
    \label{fig:angulardistribution}
\end{figure}

\bigskip
\noindent{\bf Dark photon angular and energy distributions}: In Fig.~\ref{fig:angulardistribution}, we show the unit-normalized angular distributions of dark photons from $\pi^0$ decays and $\pi^-$ absorption. 
We use GEANT4 10.5 with the FTFP\_BERT library~\cite{Agostinelli:2002hh} to simulate the case of $\pi^0$ decays. 
On the other hand, the produced monochromatic photon via the $\pi^-$ absorption process induces nuclear fragmentation processes, eventually resulting in a bunch of soft photons ($\lesssim 20$ MeV). 
We take into account such photons in our data analysis, although their contribution to our signal sensitivities is almost negligible since the nuclear $E_r$ produced by the DM emerging from the decays of dark photon produced by these soft photons is about 1 keV which is outside the range $14<E_r<26$ keV considered for the events for the DM analysis. 

Further, once a dark photon is monochromatically created in the $\pi^-$ absorption process instead of a photon, it hardly participates in further nuclear disintegration processes because its interaction with nucleons inside the mercury atom is very weak/negligible due to the kinetic mixing suppression. 

$\pi^-$ may get involved in the charge exchange process, $\pi^- + p \to \pi^0 + n$, as mentioned in the main text. Our GEANT4 simulation suggests that most of $\pi^-$ do not result in $\pi^0$, so we assume that the absorption process happens to them.
Therefore, we develop our own simulation code to take care of these single-energy-valued dark photons from the $\pi^-$ absorption case. However, $\pi^0$ decays provides the dominant contribution compared to the $\pi^-$ absorption for the dark photon parameter space limits  since we have about $2$ times more $\pi^0$ compared to $\pi^-$ per proton on target.
The dark photon is emitted isotropically in this $\pi^{-}$ absorption process whereas the dark photons emerging from the  $\pi^0$ decay have a relatively large contribution in the forward direction. 

\begin{figure}[t]
    \includegraphics[width=8.4cm]{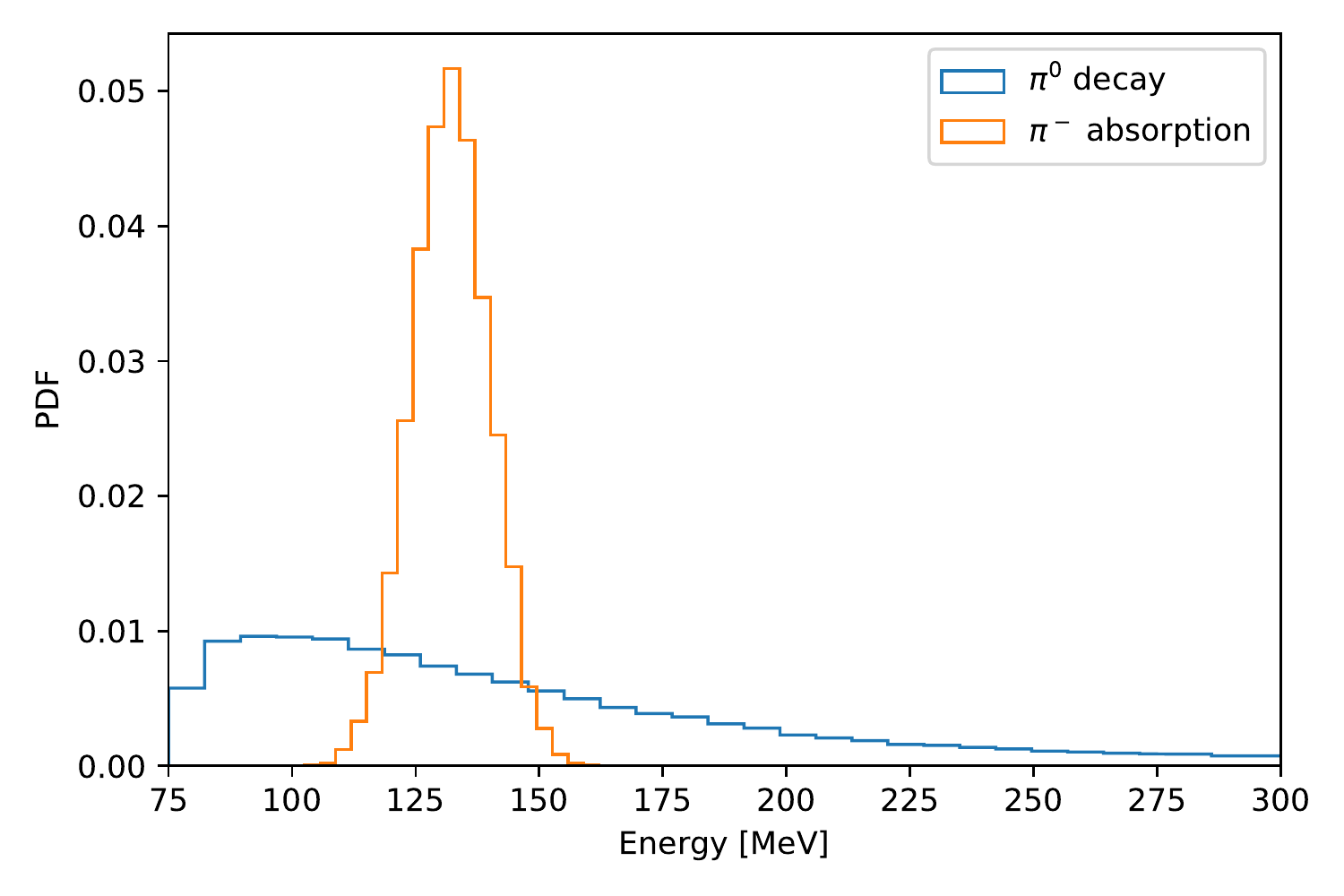}
    \includegraphics[width=8.4cm]{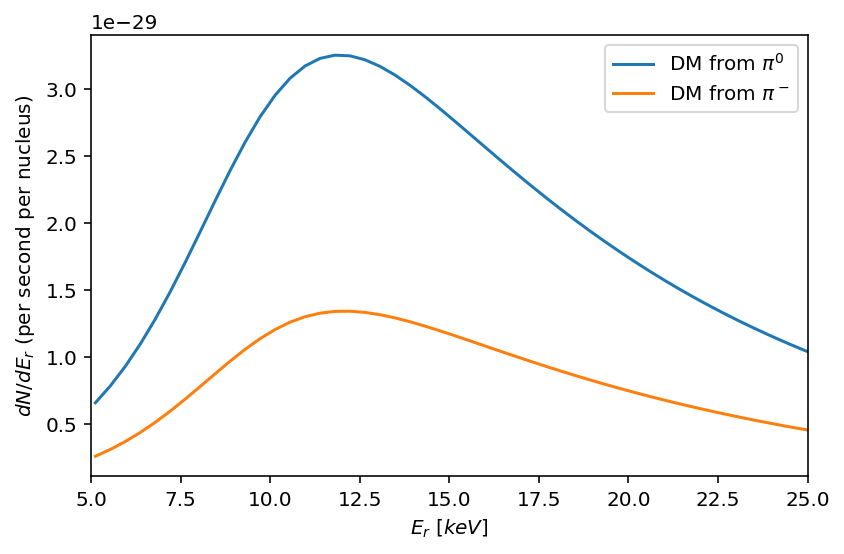}
    \caption{Top: Unit-normalized energy distributions of dark photons emerging from $\pi^-$ absorption (orange) and $\pi^0$ decays (blue). Bottom: $dN/{dE_r}$ for the DM events at the detector for $\pi^0$ decays (blue) and $\pi^-$ absorption (orange).
    }
    \label{fig:energy}
\end{figure}

In the top panel of Fig.~\ref{fig:energy}, we show the unit-normalized dark photon energy spectra from $\pi^0$ decays and $\pi^-$ absorption.  
By contrast, in the bottom panel of Fig.~\ref{fig:energy}, we show the number of DM events per second per nucleus from the $\pi^0$ decays and the $\pi^-$ absorption process where the former  surpasses the latter by at least a factor of 2 at the detector.

\bibliography{main}

\begin{thebibliography}{49}
\expandafter\ifx\csname natexlab\endcsname\relax\def\natexlab#1{#1}\fi
\expandafter\ifx\csname bibnamefont\endcsname\relax
  \def\bibnamefont#1{#1}\fi
\expandafter\ifx\csname bibfnamefont\endcsname\relax
  \def\bibfnamefont#1{#1}\fi
\expandafter\ifx\csname citenamefont\endcsname\relax
  \def\citenamefont#1{#1}\fi
\expandafter\ifx\csname url\endcsname\relax
  \def\url#1{\texttt{#1}}\fi
\expandafter\ifx\csname urlprefix\endcsname\relax\def\urlprefix{URL }\fi
\providecommand{\bibinfo}[2]{#2}
\providecommand{\eprint}[2][]{\url{#2}}

\bibitem[{\citenamefont{Aprile et~al.}(2018)}]{Aprile:2018dbl}
\bibinfo{author}{\bibfnamefont{E.}~\bibnamefont{Aprile}} \bibnamefont{et~al.}
  (\bibinfo{collaboration}{XENON}), \bibinfo{journal}{Phys. Rev. Lett.}
  \textbf{\bibinfo{volume}{121}}, \bibinfo{pages}{111302}
  (\bibinfo{year}{2018}), \eprint{1805.12562}.

\bibitem[{\citenamefont{Battaglieri et~al.}(2017)}]{Battaglieri:2017aum}
\bibinfo{author}{\bibfnamefont{M.}~\bibnamefont{Battaglieri}}
  \bibnamefont{et~al.}, in \emph{\bibinfo{booktitle}{{U.S. Cosmic Visions: New
  Ideas in Dark Matter College Park, MD, USA, March 23-25, 2017}}}
  (\bibinfo{year}{2017}), \eprint{1707.04591},
  \urlprefix\url{http://lss.fnal.gov/archive/2017/conf/fermilab-conf-17-282-ae-ppd-t.pdf}.

\bibitem[{\citenamefont{Huh et~al.}(2008)\citenamefont{Huh, Kim, Park, and
  Park}}]{Huh:2007zw}
\bibinfo{author}{\bibfnamefont{J.-H.} \bibnamefont{Huh}},
  \bibinfo{author}{\bibfnamefont{J.~E.} \bibnamefont{Kim}},
  \bibinfo{author}{\bibfnamefont{J.-C.} \bibnamefont{Park}}, \bibnamefont{and}
  \bibinfo{author}{\bibfnamefont{S.~C.} \bibnamefont{Park}},
  \bibinfo{journal}{Phys. Rev.} \textbf{\bibinfo{volume}{D77}},
  \bibinfo{pages}{123503} (\bibinfo{year}{2008}), \eprint{0711.3528}.

\bibitem[{\citenamefont{Pospelov et~al.}(2008)\citenamefont{Pospelov, Ritz, and
  Voloshin}}]{Pospelov:2007mp}
\bibinfo{author}{\bibfnamefont{M.}~\bibnamefont{Pospelov}},
  \bibinfo{author}{\bibfnamefont{A.}~\bibnamefont{Ritz}}, \bibnamefont{and}
  \bibinfo{author}{\bibfnamefont{M.~B.} \bibnamefont{Voloshin}},
  \bibinfo{journal}{Phys. Lett.} \textbf{\bibinfo{volume}{B662}},
  \bibinfo{pages}{53} (\bibinfo{year}{2008}), \eprint{0711.4866}.

\bibitem[{\citenamefont{Hooper and Zurek}(2008)}]{Hooper:2008im}
\bibinfo{author}{\bibfnamefont{D.}~\bibnamefont{Hooper}} \bibnamefont{and}
  \bibinfo{author}{\bibfnamefont{K.~M.} \bibnamefont{Zurek}},
  \bibinfo{journal}{Phys. Rev.} \textbf{\bibinfo{volume}{D77}},
  \bibinfo{pages}{087302} (\bibinfo{year}{2008}), \eprint{0801.3686}.

\bibitem[{\citenamefont{Cheung et~al.}(2009)\citenamefont{Cheung, Ruderman,
  Wang, and Yavin}}]{Cheung:2009qd}
\bibinfo{author}{\bibfnamefont{C.}~\bibnamefont{Cheung}},
  \bibinfo{author}{\bibfnamefont{J.~T.} \bibnamefont{Ruderman}},
  \bibinfo{author}{\bibfnamefont{L.-T.} \bibnamefont{Wang}}, \bibnamefont{and}
  \bibinfo{author}{\bibfnamefont{I.}~\bibnamefont{Yavin}},
  \bibinfo{journal}{Phys. Rev.} \textbf{\bibinfo{volume}{D80}},
  \bibinfo{pages}{035008} (\bibinfo{year}{2009}), \eprint{0902.3246}.

\bibitem[{\citenamefont{Essig et~al.}(2010)\citenamefont{Essig, Kaplan,
  Schuster, and Toro}}]{Essig:2010ye}
\bibinfo{author}{\bibfnamefont{R.}~\bibnamefont{Essig}},
  \bibinfo{author}{\bibfnamefont{J.}~\bibnamefont{Kaplan}},
  \bibinfo{author}{\bibfnamefont{P.}~\bibnamefont{Schuster}}, \bibnamefont{and}
  \bibinfo{author}{\bibfnamefont{N.}~\bibnamefont{Toro}},
  \bibinfo{journal}{Submitted to: Physical Review D}  (\bibinfo{year}{2010}),
  \eprint{1004.0691}.

\bibitem[{\citenamefont{Essig et~al.}(2013{\natexlab{a}})}]{Essig:2013lka}
\bibinfo{author}{\bibfnamefont{R.}~\bibnamefont{Essig}} \bibnamefont{et~al.},
  in \emph{\bibinfo{booktitle}{{Proceedings, 2013 Community Summer Study on the
  Future of U.S. Particle Physics: Snowmass on the Mississippi (CSS2013):
  Minneapolis, MN, USA, July 29-August 6, 2013}}}
  (\bibinfo{year}{2013}{\natexlab{a}}), \eprint{1311.0029},
  \urlprefix\url{http://www.slac.stanford.edu/econf/C1307292/docs/IntensityFrontier/NewLight-17.pdf}.

\bibitem[{\citenamefont{Dutta et~al.}(2019{\natexlab{a}})\citenamefont{Dutta,
  Ghosh, and Kumar}}]{Dutta:2019fxn}
\bibinfo{author}{\bibfnamefont{B.}~\bibnamefont{Dutta}},
  \bibinfo{author}{\bibfnamefont{S.}~\bibnamefont{Ghosh}}, \bibnamefont{and}
  \bibinfo{author}{\bibfnamefont{J.}~\bibnamefont{Kumar}},
  \bibinfo{journal}{Phys. Rev.} \textbf{\bibinfo{volume}{D100}},
  \bibinfo{pages}{075028} (\bibinfo{year}{2019}{\natexlab{a}}),
  \eprint{1905.02692}.

\bibitem[{\citenamefont{Akimov et~al.}(2018{\natexlab{a}})}]{Akimov:2018ghi}
\bibinfo{author}{\bibfnamefont{D.}~\bibnamefont{Akimov}} \bibnamefont{et~al.}
  (\bibinfo{collaboration}{COHERENT}) (\bibinfo{year}{2018}{\natexlab{a}}),
  \eprint{1803.09183}.

\bibitem[{\citenamefont{Ohlsson}(2013)}]{Ohlsson:2012kf}
\bibinfo{author}{\bibfnamefont{T.}~\bibnamefont{Ohlsson}},
  \bibinfo{journal}{Rept. Prog. Phys.} \textbf{\bibinfo{volume}{76}},
  \bibinfo{pages}{044201} (\bibinfo{year}{2013}), \eprint{1209.2710}.

\bibitem[{\citenamefont{Miranda and Nunokawa}(2015)}]{Miranda:2015dra}
\bibinfo{author}{\bibfnamefont{O.~G.} \bibnamefont{Miranda}} \bibnamefont{and}
  \bibinfo{author}{\bibfnamefont{H.}~\bibnamefont{Nunokawa}},
  \bibinfo{journal}{New J. Phys.} \textbf{\bibinfo{volume}{17}},
  \bibinfo{pages}{095002} (\bibinfo{year}{2015}), \eprint{1505.06254}.

\bibitem[{\citenamefont{Coloma et~al.}(2017{\natexlab{a}})\citenamefont{Coloma,
  Denton, Gonzalez-Garcia, Maltoni, and Schwetz}}]{Coloma:2017egw}
\bibinfo{author}{\bibfnamefont{P.}~\bibnamefont{Coloma}},
  \bibinfo{author}{\bibfnamefont{P.~B.} \bibnamefont{Denton}},
  \bibinfo{author}{\bibfnamefont{M.~C.} \bibnamefont{Gonzalez-Garcia}},
  \bibinfo{author}{\bibfnamefont{M.}~\bibnamefont{Maltoni}}, \bibnamefont{and}
  \bibinfo{author}{\bibfnamefont{T.}~\bibnamefont{Schwetz}},
  \bibinfo{journal}{JHEP} \textbf{\bibinfo{volume}{04}}, \bibinfo{pages}{116}
  (\bibinfo{year}{2017}{\natexlab{a}}), \eprint{1701.04828}.

\bibitem[{\citenamefont{Coloma et~al.}(2017{\natexlab{b}})\citenamefont{Coloma,
  Gonzalez-Garcia, Maltoni, and Schwetz}}]{Coloma:2017ncl}
\bibinfo{author}{\bibfnamefont{P.}~\bibnamefont{Coloma}},
  \bibinfo{author}{\bibfnamefont{M.~C.} \bibnamefont{Gonzalez-Garcia}},
  \bibinfo{author}{\bibfnamefont{M.}~\bibnamefont{Maltoni}}, \bibnamefont{and}
  \bibinfo{author}{\bibfnamefont{T.}~\bibnamefont{Schwetz}},
  \bibinfo{journal}{Phys. Rev.} \textbf{\bibinfo{volume}{D96}},
  \bibinfo{pages}{115007} (\bibinfo{year}{2017}{\natexlab{b}}),
  \eprint{1708.02899}.

\bibitem[{\citenamefont{Liao and Marfatia}(2017)}]{Liao:2017uzy}
\bibinfo{author}{\bibfnamefont{J.}~\bibnamefont{Liao}} \bibnamefont{and}
  \bibinfo{author}{\bibfnamefont{D.}~\bibnamefont{Marfatia}},
  \bibinfo{journal}{Phys. Lett.} \textbf{\bibinfo{volume}{B775}},
  \bibinfo{pages}{54} (\bibinfo{year}{2017}), \eprint{1708.04255}.

\bibitem[{\citenamefont{Dent et~al.}(2018)\citenamefont{Dent, Dutta, Liao,
  Newstead, Strigari, and Walker}}]{Dent:2017mpr}
\bibinfo{author}{\bibfnamefont{J.~B.} \bibnamefont{Dent}},
  \bibinfo{author}{\bibfnamefont{B.}~\bibnamefont{Dutta}},
  \bibinfo{author}{\bibfnamefont{S.}~\bibnamefont{Liao}},
  \bibinfo{author}{\bibfnamefont{J.~L.} \bibnamefont{Newstead}},
  \bibinfo{author}{\bibfnamefont{L.~E.} \bibnamefont{Strigari}},
  \bibnamefont{and} \bibinfo{author}{\bibfnamefont{J.~W.}
  \bibnamefont{Walker}}, \bibinfo{journal}{Phys. Rev.}
  \textbf{\bibinfo{volume}{D97}}, \bibinfo{pages}{035009}
  (\bibinfo{year}{2018}), \eprint{1711.03521}.

\bibitem[{\citenamefont{Billard et~al.}(2018)\citenamefont{Billard, Johnston,
  and Kavanagh}}]{Billard:2018jnl}
\bibinfo{author}{\bibfnamefont{J.}~\bibnamefont{Billard}},
  \bibinfo{author}{\bibfnamefont{J.}~\bibnamefont{Johnston}}, \bibnamefont{and}
  \bibinfo{author}{\bibfnamefont{B.~J.} \bibnamefont{Kavanagh}},
  \bibinfo{journal}{JCAP} \textbf{\bibinfo{volume}{1811}}, \bibinfo{pages}{016}
  (\bibinfo{year}{2018}), \eprint{1805.01798}.

\bibitem[{\citenamefont{Lindner et~al.}(2017)\citenamefont{Lindner, Rodejohann,
  and Xu}}]{Lindner:2016wff}
\bibinfo{author}{\bibfnamefont{M.}~\bibnamefont{Lindner}},
  \bibinfo{author}{\bibfnamefont{W.}~\bibnamefont{Rodejohann}},
  \bibnamefont{and} \bibinfo{author}{\bibfnamefont{X.-J.} \bibnamefont{Xu}},
  \bibinfo{journal}{JHEP} \textbf{\bibinfo{volume}{03}}, \bibinfo{pages}{097}
  (\bibinfo{year}{2017}), \eprint{1612.04150}.

\bibitem[{\citenamefont{Farzan et~al.}(2018)\citenamefont{Farzan, Lindner,
  Rodejohann, and Xu}}]{Farzan:2018gtr}
\bibinfo{author}{\bibfnamefont{Y.}~\bibnamefont{Farzan}},
  \bibinfo{author}{\bibfnamefont{M.}~\bibnamefont{Lindner}},
  \bibinfo{author}{\bibfnamefont{W.}~\bibnamefont{Rodejohann}},
  \bibnamefont{and} \bibinfo{author}{\bibfnamefont{X.-J.} \bibnamefont{Xu}},
  \bibinfo{journal}{JHEP} \textbf{\bibinfo{volume}{05}}, \bibinfo{pages}{066}
  (\bibinfo{year}{2018}), \eprint{1802.05171}.

\bibitem[{\citenamefont{Brdar et~al.}(2018)\citenamefont{Brdar, Rodejohann, and
  Xu}}]{Brdar:2018qqj}
\bibinfo{author}{\bibfnamefont{V.}~\bibnamefont{Brdar}},
  \bibinfo{author}{\bibfnamefont{W.}~\bibnamefont{Rodejohann}},
  \bibnamefont{and} \bibinfo{author}{\bibfnamefont{X.-J.} \bibnamefont{Xu}},
  \bibinfo{journal}{JHEP} \textbf{\bibinfo{volume}{12}}, \bibinfo{pages}{024}
  (\bibinfo{year}{2018}), \eprint{1810.03626}.

\bibitem[{\citenamefont{Aristizabal~Sierra
  et~al.}(2018)\citenamefont{Aristizabal~Sierra, De~Romeri, and
  Rojas}}]{AristizabalSierra:2018eqm}
\bibinfo{author}{\bibfnamefont{D.}~\bibnamefont{Aristizabal~Sierra}},
  \bibinfo{author}{\bibfnamefont{V.}~\bibnamefont{De~Romeri}},
  \bibnamefont{and} \bibinfo{author}{\bibfnamefont{N.}~\bibnamefont{Rojas}},
  \bibinfo{journal}{Phys. Rev.} \textbf{\bibinfo{volume}{D98}},
  \bibinfo{pages}{075018} (\bibinfo{year}{2018}), \eprint{1806.07424}.

\bibitem[{\citenamefont{Datta et~al.}(2019)\citenamefont{Datta, Dutta, Liao,
  Marfatia, and Strigari}}]{Datta:2018xty}
\bibinfo{author}{\bibfnamefont{A.}~\bibnamefont{Datta}},
  \bibinfo{author}{\bibfnamefont{B.}~\bibnamefont{Dutta}},
  \bibinfo{author}{\bibfnamefont{S.}~\bibnamefont{Liao}},
  \bibinfo{author}{\bibfnamefont{D.}~\bibnamefont{Marfatia}}, \bibnamefont{and}
  \bibinfo{author}{\bibfnamefont{L.~E.} \bibnamefont{Strigari}},
  \bibinfo{journal}{JHEP} \textbf{\bibinfo{volume}{01}}, \bibinfo{pages}{091}
  (\bibinfo{year}{2019}), \eprint{1808.02611}.

\bibitem[{\citenamefont{Kosmas et~al.}(2017)\citenamefont{Kosmas, Papoulias,
  Tortola, and Valle}}]{Kosmas:2017zbh}
\bibinfo{author}{\bibfnamefont{T.~S.} \bibnamefont{Kosmas}},
  \bibinfo{author}{\bibfnamefont{D.~K.} \bibnamefont{Papoulias}},
  \bibinfo{author}{\bibfnamefont{M.}~\bibnamefont{Tortola}}, \bibnamefont{and}
  \bibinfo{author}{\bibfnamefont{J.~W.~F.} \bibnamefont{Valle}},
  \bibinfo{journal}{Phys. Rev.} \textbf{\bibinfo{volume}{D96}},
  \bibinfo{pages}{063013} (\bibinfo{year}{2017}), \eprint{1703.00054}.

\bibitem[{\citenamefont{Blanco et~al.}(2019)\citenamefont{Blanco, Hooper, and
  Machado}}]{Blanco:2019vyp}
\bibinfo{author}{\bibfnamefont{C.}~\bibnamefont{Blanco}},
  \bibinfo{author}{\bibfnamefont{D.}~\bibnamefont{Hooper}}, \bibnamefont{and}
  \bibinfo{author}{\bibfnamefont{P.}~\bibnamefont{Machado}}
  (\bibinfo{year}{2019}), \eprint{1901.08094}.

\bibitem[{\citenamefont{Ciuffoli et~al.}(2018)\citenamefont{Ciuffoli, Evslin,
  Fu, and Tang}}]{Ciuffoli:2018qem}
\bibinfo{author}{\bibfnamefont{E.}~\bibnamefont{Ciuffoli}},
  \bibinfo{author}{\bibfnamefont{J.}~\bibnamefont{Evslin}},
  \bibinfo{author}{\bibfnamefont{Q.}~\bibnamefont{Fu}}, \bibnamefont{and}
  \bibinfo{author}{\bibfnamefont{J.}~\bibnamefont{Tang}},
  \bibinfo{journal}{Phys. Rev.} \textbf{\bibinfo{volume}{D97}},
  \bibinfo{pages}{113003} (\bibinfo{year}{2018}), \eprint{1801.02166}.

\bibitem[{\citenamefont{Aristizabal~Sierra
  et~al.}(2019)\citenamefont{Aristizabal~Sierra, Liao, and
  Marfatia}}]{AristizabalSierra:2019zmy}
\bibinfo{author}{\bibfnamefont{D.}~\bibnamefont{Aristizabal~Sierra}},
  \bibinfo{author}{\bibfnamefont{J.}~\bibnamefont{Liao}}, \bibnamefont{and}
  \bibinfo{author}{\bibfnamefont{D.}~\bibnamefont{Marfatia}}
  (\bibinfo{year}{2019}), \eprint{1902.07398}.

\bibitem[{\citenamefont{Papoulias et~al.}(2019)\citenamefont{Papoulias, Kosmas,
  Sahu, Kota, and Hota}}]{Papoulias:2019lfi}
\bibinfo{author}{\bibfnamefont{D.~K.} \bibnamefont{Papoulias}},
  \bibinfo{author}{\bibfnamefont{T.~S.} \bibnamefont{Kosmas}},
  \bibinfo{author}{\bibfnamefont{R.}~\bibnamefont{Sahu}},
  \bibinfo{author}{\bibfnamefont{V.~K.~B.} \bibnamefont{Kota}},
  \bibnamefont{and} \bibinfo{author}{\bibfnamefont{M.}~\bibnamefont{Hota}}
  (\bibinfo{year}{2019}), \eprint{1903.03722}.

\bibitem[{\citenamefont{Dutta et~al.}(2019{\natexlab{b}})\citenamefont{Dutta,
  Liao, Sinha, and Strigari}}]{Dutta:2019eml}
\bibinfo{author}{\bibfnamefont{B.}~\bibnamefont{Dutta}},
  \bibinfo{author}{\bibfnamefont{S.}~\bibnamefont{Liao}},
  \bibinfo{author}{\bibfnamefont{S.}~\bibnamefont{Sinha}}, \bibnamefont{and}
  \bibinfo{author}{\bibfnamefont{L.~E.} \bibnamefont{Strigari}}
  (\bibinfo{year}{2019}{\natexlab{b}}), \eprint{1903.10666}.

\bibitem[{\citenamefont{deNiverville et~al.}(2015)\citenamefont{deNiverville,
  Pospelov, and Ritz}}]{deNiverville:2015mwa}
\bibinfo{author}{\bibfnamefont{P.}~\bibnamefont{deNiverville}},
  \bibinfo{author}{\bibfnamefont{M.}~\bibnamefont{Pospelov}}, \bibnamefont{and}
  \bibinfo{author}{\bibfnamefont{A.}~\bibnamefont{Ritz}},
  \bibinfo{journal}{Phys. Rev.} \textbf{\bibinfo{volume}{D92}},
  \bibinfo{pages}{095005} (\bibinfo{year}{2015}), \eprint{1505.07805}.

\bibitem[{\citenamefont{Ge and Shoemaker}(2018)}]{Ge:2017mcq}
\bibinfo{author}{\bibfnamefont{S.-F.} \bibnamefont{Ge}} \bibnamefont{and}
  \bibinfo{author}{\bibfnamefont{I.~M.} \bibnamefont{Shoemaker}},
  \bibinfo{journal}{JHEP} \textbf{\bibinfo{volume}{11}}, \bibinfo{pages}{066}
  (\bibinfo{year}{2018}), \eprint{1710.10889}.

\bibitem[{\citenamefont{Akimov et~al.}(2018{\natexlab{b}})}]{akimov}
\bibinfo{author}{\bibfnamefont{D.}~\bibnamefont{Akimov}} \bibnamefont{et~al.}
  (\bibinfo{collaboration}{COHERENT}) (\bibinfo{year}{2018}{\natexlab{b}}),
  \eprint{1804.09459}.

\bibitem[{Coh(Rebeca Rapp)}]{Coherent}
\bibinfo{journal}{Private communication with Coherent Collaboration}
  (\bibinfo{year}{Rebeca Rapp}).

\bibitem[{\citenamefont{Holdom}(1986)}]{Holdom:1985ag}
\bibinfo{author}{\bibfnamefont{B.}~\bibnamefont{Holdom}},
  \bibinfo{journal}{Phys. Lett.} \textbf{\bibinfo{volume}{166B}},
  \bibinfo{pages}{196} (\bibinfo{year}{1986}).

\bibitem[{\citenamefont{del Aguila et~al.}(1988)\citenamefont{del Aguila,
  Coughlan, and Quiros}}]{delAguila:1988jz}
\bibinfo{author}{\bibfnamefont{F.}~\bibnamefont{del Aguila}},
  \bibinfo{author}{\bibfnamefont{G.~D.} \bibnamefont{Coughlan}},
  \bibnamefont{and} \bibinfo{author}{\bibfnamefont{M.}~\bibnamefont{Quiros}},
  \bibinfo{journal}{Nucl. Phys.} \textbf{\bibinfo{volume}{B307}},
  \bibinfo{pages}{633} (\bibinfo{year}{1988}), \bibinfo{note}{[Erratum: Nucl.
  Phys.B312,751(1989)]}.

\bibitem[{\citenamefont{Babu et~al.}(1998)\citenamefont{Babu, Kolda, and
  March-Russell}}]{Babu:1997st}
\bibinfo{author}{\bibfnamefont{K.~S.} \bibnamefont{Babu}},
  \bibinfo{author}{\bibfnamefont{C.~F.} \bibnamefont{Kolda}}, \bibnamefont{and}
  \bibinfo{author}{\bibfnamefont{J.}~\bibnamefont{March-Russell}},
  \bibinfo{journal}{Phys. Rev.} \textbf{\bibinfo{volume}{D57}},
  \bibinfo{pages}{6788} (\bibinfo{year}{1998}), \eprint{hep-ph/9710441}.

\bibitem[{\citenamefont{Agostinelli et~al.}(2003)}]{Agostinelli:2002hh}
\bibinfo{author}{\bibfnamefont{S.}~\bibnamefont{Agostinelli}}
  \bibnamefont{et~al.} (\bibinfo{collaboration}{GEANT4}),
  \bibinfo{journal}{Nucl. Instrum. Meth.} \textbf{\bibinfo{volume}{A506}},
  \bibinfo{pages}{250} (\bibinfo{year}{2003}).

\bibitem[{\citenamefont{Cadeddu et~al.}(2018)\citenamefont{Cadeddu, Giunti, Li,
  and Zhang}}]{Cadeddu:2017etk}
\bibinfo{author}{\bibfnamefont{M.}~\bibnamefont{Cadeddu}},
  \bibinfo{author}{\bibfnamefont{C.}~\bibnamefont{Giunti}},
  \bibinfo{author}{\bibfnamefont{Y.~F.} \bibnamefont{Li}}, \bibnamefont{and}
  \bibinfo{author}{\bibfnamefont{Y.~Y.} \bibnamefont{Zhang}},
  \bibinfo{journal}{Phys. Rev. Lett.} \textbf{\bibinfo{volume}{120}},
  \bibinfo{pages}{072501} (\bibinfo{year}{2018}), \eprint{1710.02730}.

\bibitem[{\citenamefont{Scholberg}(2018)}]{Scholberg:2018vwg}
\bibinfo{author}{\bibfnamefont{K.}~\bibnamefont{Scholberg}}
  (\bibinfo{collaboration}{COHERENT}), \bibinfo{journal}{PoS}
  \textbf{\bibinfo{volume}{NuFact2017}}, \bibinfo{pages}{020}
  (\bibinfo{year}{2018}), \eprint{1801.05546}.

\bibitem[{\citenamefont{Akimov et~al.}(2017)}]{Akimov:2017ade}
\bibinfo{author}{\bibfnamefont{D.}~\bibnamefont{Akimov}} \bibnamefont{et~al.}
  (\bibinfo{collaboration}{COHERENT}), \bibinfo{journal}{Science}
  \textbf{\bibinfo{volume}{357}}, \bibinfo{pages}{1123} (\bibinfo{year}{2017}),
  \eprint{1708.01294}.

\bibitem[{\citenamefont{Tanabashi et~al.}(2018)}]{Tanabashi:2018oca}
\bibinfo{author}{\bibfnamefont{M.}~\bibnamefont{Tanabashi}}
  \bibnamefont{et~al.} (\bibinfo{collaboration}{Particle Data Group}),
  \bibinfo{journal}{Phys. Rev.} \textbf{\bibinfo{volume}{D98}},
  \bibinfo{pages}{030001} (\bibinfo{year}{2018}).

\bibitem[{\citenamefont{Essig et~al.}(2013{\natexlab{b}})\citenamefont{Essig,
  Mardon, Papucci, Volansky, and Zhong}}]{Essig:2013vha}
\bibinfo{author}{\bibfnamefont{R.}~\bibnamefont{Essig}},
  \bibinfo{author}{\bibfnamefont{J.}~\bibnamefont{Mardon}},
  \bibinfo{author}{\bibfnamefont{M.}~\bibnamefont{Papucci}},
  \bibinfo{author}{\bibfnamefont{T.}~\bibnamefont{Volansky}}, \bibnamefont{and}
  \bibinfo{author}{\bibfnamefont{Y.-M.} \bibnamefont{Zhong}},
  \bibinfo{journal}{JHEP} \textbf{\bibinfo{volume}{11}}, \bibinfo{pages}{167}
  (\bibinfo{year}{2013}{\natexlab{b}}), \eprint{1309.5084}.

\bibitem[{\citenamefont{Dror et~al.}(2017)\citenamefont{Dror, Lasenby, and
  Pospelov}}]{Dror:2017nsg}
\bibinfo{author}{\bibfnamefont{J.~A.} \bibnamefont{Dror}},
  \bibinfo{author}{\bibfnamefont{R.}~\bibnamefont{Lasenby}}, \bibnamefont{and}
  \bibinfo{author}{\bibfnamefont{M.}~\bibnamefont{Pospelov}},
  \bibinfo{journal}{Phys. Rev.} \textbf{\bibinfo{volume}{D96}},
  \bibinfo{pages}{075036} (\bibinfo{year}{2017}), \eprint{1707.01503}.

\bibitem[{\citenamefont{Artamonov et~al.}(2009)}]{Artamonov:2009sz}
\bibinfo{author}{\bibfnamefont{A.~V.} \bibnamefont{Artamonov}}
  \bibnamefont{et~al.} (\bibinfo{collaboration}{BNL-E949}),
  \bibinfo{journal}{Phys. Rev.} \textbf{\bibinfo{volume}{D79}},
  \bibinfo{pages}{092004} (\bibinfo{year}{2009}), \eprint{0903.0030}.

\bibitem[{\citenamefont{Davoudiasl et~al.}(2014)\citenamefont{Davoudiasl, Lee,
  and Marciano}}]{Davoudiasl:2014kua}
\bibinfo{author}{\bibfnamefont{H.}~\bibnamefont{Davoudiasl}},
  \bibinfo{author}{\bibfnamefont{H.-S.} \bibnamefont{Lee}}, \bibnamefont{and}
  \bibinfo{author}{\bibfnamefont{W.~J.} \bibnamefont{Marciano}},
  \bibinfo{journal}{Phys. Rev.} \textbf{\bibinfo{volume}{D89}},
  \bibinfo{pages}{095006} (\bibinfo{year}{2014}), \eprint{1402.3620}.

\bibitem[{\citenamefont{Åkesson et~al.}(2018)}]{Akesson:2018vlm}
\bibinfo{author}{\bibfnamefont{T.}~\bibnamefont{Åkesson}} \bibnamefont{et~al.}
  (\bibinfo{collaboration}{LDMX}) (\bibinfo{year}{2018}), \eprint{1808.05219}.

\bibitem[{\citenamefont{Banerjee et~al.}(2019)}]{NA64:2019imj}
\bibinfo{author}{\bibfnamefont{D.}~\bibnamefont{Banerjee}}
  \bibnamefont{et~al.}, \bibinfo{journal}{Phys. Rev. Lett.}
  \textbf{\bibinfo{volume}{123}}, \bibinfo{pages}{121801}
  (\bibinfo{year}{2019}), \eprint{1906.00176}.

\bibitem[{\citenamefont{De~Romeri et~al.}(2019)\citenamefont{De~Romeri, Kelly,
  and Machado}}]{DeRomeri:2019kic}
\bibinfo{author}{\bibfnamefont{V.}~\bibnamefont{De~Romeri}},
  \bibinfo{author}{\bibfnamefont{K.~J.} \bibnamefont{Kelly}}, \bibnamefont{and}
  \bibinfo{author}{\bibfnamefont{P.~A.~N.} \bibnamefont{Machado}}
  (\bibinfo{year}{2019}), \eprint{1903.10505}.

\bibitem[{\citenamefont{Ajimura et~al.}(2017)}]{Ajimura:2017fld}
\bibinfo{author}{\bibfnamefont{S.}~\bibnamefont{Ajimura}} \bibnamefont{et~al.}
  (\bibinfo{year}{2017}), \eprint{1705.08629}.

\bibitem[{\citenamefont{Aguilar-Arevalo et~al.}(2018)}]{CCM}
\bibinfo{author}{\bibfnamefont{A.~A.} \bibnamefont{Aguilar-Arevalo}}
  \bibnamefont{et~al.} (\bibinfo{year}{2018}).

\end{thebibliography}

\end{document}